\documentstyle[sprocl,epsfig]{article}

\def\ab{\bar{\alpha}}
\def\be{\begin{equation}}
\def\ee{\end{equation}}
\def\bea{\begin{eqnarray}}
\def\eea{\end{eqnarray}}

\newcommand{\ltsima}{$\; \buildrel < \over \sim \;$}
\newcommand{\simlt}{\lower.5ex\hbox{\ltsima}}
\newcommand{\gtsima}{$\; \buildrel > \over \sim \;$}
\newcommand{\simgt}{\lower.5ex\hbox{\gtsima}}

\newcommand{\cgs}{${\rm erg~cm}^{-2}~{\rm s}^{-1}$} 

\newcommand{\pn}{\par\noindent}

\def\lesssim{\mathrel{\hbox{\rlap{\hbox{\lower4pt\hbox{$\sim$}}}\hbox{$<$}}}}
\def\gtrsim{\mathrel{\hbox{\rlap{\hbox{\lower4pt\hbox{$\sim$}}}\hbox{$>$}}}}

\def\aox{$\alpha_{\rm ox}$}

\def\ab1450{$AB_{1450(1+z)}$}

\def\chandra{{\it Chandra\/}}

\def\heao1{{\it HEAO-1\/}}

\def\rosat{{\it ROSAT\/}}

\def\xmm{{XMM-{\it Newton\/}}}

\def \sait #1 #2 {{\em Mem.\ Soc.\ Astron.\ It.\/} {\bf #1}, #2\ }
\def \mess #1 #2 {{\em The Messenger\/} {\bf #1}, #2\ }
\def \astrnach #1 #2 {{\em Astron. Nach.\/} {\bf #1}, #2\ }
\def \aap #1 #2 {{\em Astron. Astrophys.\/} {\bf #1}, #2\ }
\def \aal #1 #2 {{\em Astron. Astrophys. Lett.\/} {\bf #1}, L#2\ }
\def \aar #1 #2 {{\em Astron. Astrophys. Rev.\/} {\bf #1}, #2\ }
\def \aas #1 #2 {{\em Astron. Astrophys. Suppl. Ser.\/} {\bf #1}, #2\ }
\def \aj #1 #2 {{\em Astron. J.\/} {\bf #1}, #2\ }
\def \araa #1 #2 {{\em Ann. Rev. Astron. Astrophys.\/} {\bf #1}, #2\ }
\def \apj #1 #2 {{\em Astrophys. J.\/} {\bf #1}, #2\ }
\def \apjl #1 #2 {{\em Astrophys. J. Lett.\/} {\bf #1}, L#2\ }
\def \apjs #1 #2 {{\em Astrophys. J. Suppl.\/} {\bf #1}, #2\ }
\def \apss #1 #2 {{\em Astrophys. Space Sci.\/} {\bf #1}, #2\ }
\def \asr #1 #2 {{\em Adv. Space Res.\/} {\bf #1}, #2\ }
\def \baic #1 #2 {{\em Bull. Astron. Inst. Czechosl.\/} {\bf #1}, #2\ }
\def \jqsrt #1 #2 {{\em J. Quant. Spectrosc. Radiat. Transfer\/} {\bf #1}, #2\ }
\def \mnras #1 #2 {{\em Mon. Not. R. Astr. Soc.\/} {\bf #1}, #2\ }
\def \mem #1 #2 {{\em Mem. R. Astr. Soc.\/} {\bf #1}, #2\ }
\def \plr #1 #2 {{\em Phys. Lett. Rev.\/} {\bf #1}, #2\ }
\def \pasj #1 #2 {{\em Publ. Astron. Soc. Japan\/} {\bf #1}, #2\ }
\def \pasp #1 #2 {{\em Publ. Astr. Soc. Pacific\/} {\bf #1}, #2\ }
\def \nat #1 #2 {{\em Nature\/} {\bf #1}, #2\ }
\def \etal {{\it et~al.}}

\begin{document}

\title{X-RAY EMISSION FROM THE MOST LUMINOUS $\mathbf{Z>4}$ 
PALOMAR DIGITAL SKY SURVEY QUASARS: THE CHANDRA VIEW} 

\author{C. VIGNALI, W.~N. BRANDT, D.~P. SCHNEIDER, G.~P. GARMIRE}

\address{Department of Astronomy \& Astrophysics, 
The Pennsylvania State University, \\
525 Davey Laboratory, University Park, PA 16802, USA \\ 
E-mail: chris, niel, dps, \& garmire@astro.psu.edu}

\author{S. KASPI}

\address{School of Physics and Astronomy, Raymond and Beverly Sackler 
Faculty of Exact Sciences, Tel-Aviv University, 
Tel-Aviv 69978, Israel \\
E-mail: shai@wise.tau.ac.il}


\maketitle
\abstracts{
We present the results obtained from exploratory \chandra\ 
observations of nine high-redshift \hbox{($z$=4.09--4.51)} quasars, 
selected from among the optically brightest Palomar Digital Sky Survey 
quasars known. 
Their broad-band spectral energy distributions are characterized, on average, 
by steeper \aox\ values ($\langle\alpha_{\rm ox}\rangle$=$-$1.81$\pm{0.03}$) 
than those of lower-redshift, lower-luminosity samples of quasars. 
We find a significant correlation between \ab1450\ magnitude and 
soft X-ray flux, suggesting that the engine powering the UV and 
X-ray emission is the same. 
The joint \hbox{$\approx$~2--30~keV} rest-frame X-ray spectrum 
is well parameterized by a $\Gamma\approx2.0\pm{0.2}$ power-law model with 
no evidence for intrinsic absorption 
($N_{\rm H}\simlt8.8\times10^{21}$~cm$^{-2}$). 
}

\section{Introduction. Observational and scientific strategy}

Our understanding of quasar X-ray properties at $z>4$ has advanced 
rapidly over the past few years, mostly thanks to the capabilities of 
\chandra\ and \xmm. 
Prior to 2000 there were only six X-ray detected quasars at $z>4$. 
The first systematic X-ray study of $z>4$ quasars, carried out 
using archival \rosat\ data{\,}\cite{k00}, doubled the 
number of X-ray detected quasars at $z>4$. 
Recent ground-based optical surveys have discovered 
a large number of AGNs at $z>4$. 
In particular, the Palomar Digital Sky Survey{\,}\cite{d98} (PSS), 
the Automatic Plate Measuring facility survey{\,}\cite{i91} (BRI), and the 
Sloan Digital Sky Survey{\,}\cite{y00} (SDSS) have discovered more than 
300 $z>4$ 
quasars\footnote{See http://www.astro.caltech.edu/$\sim$george/z4.qsos 
for a listing of high-redshift quasars.} up to $z=6.28$. 
To define the basic X-ray properties of $z>4$ quasars 
in two different regions of the luminosity-redshift parameter space, 
we started a program to observe with \chandra\ and \xmm\ both the 
optically brightest \hbox{$z\approx$~4--4.6} PSS/BRI quasars and 
the higher-redshift, optically fainter SDSS quasars (Fig.~1, left panel). 
Since the pioneering work with \rosat{\,}\cite{k00}, 
the number of AGNs with \hbox{X-ray} detections has increased significantly 
to $\approx$~50 (see Fig.~1, right 
panel)\footnote{See http://www.astro.psu.edu/users/niel/papers/highz-xray-detected.dat for a regularly updated compilation of X-ray detections at $z>4$.}
in the redshift range $z\approx$~4--6.3.{\,}\cite{b01a,v01,b02,be02,v02}
\vskip -0.3cm
\begin{figure}[!h]
\psfig{figure=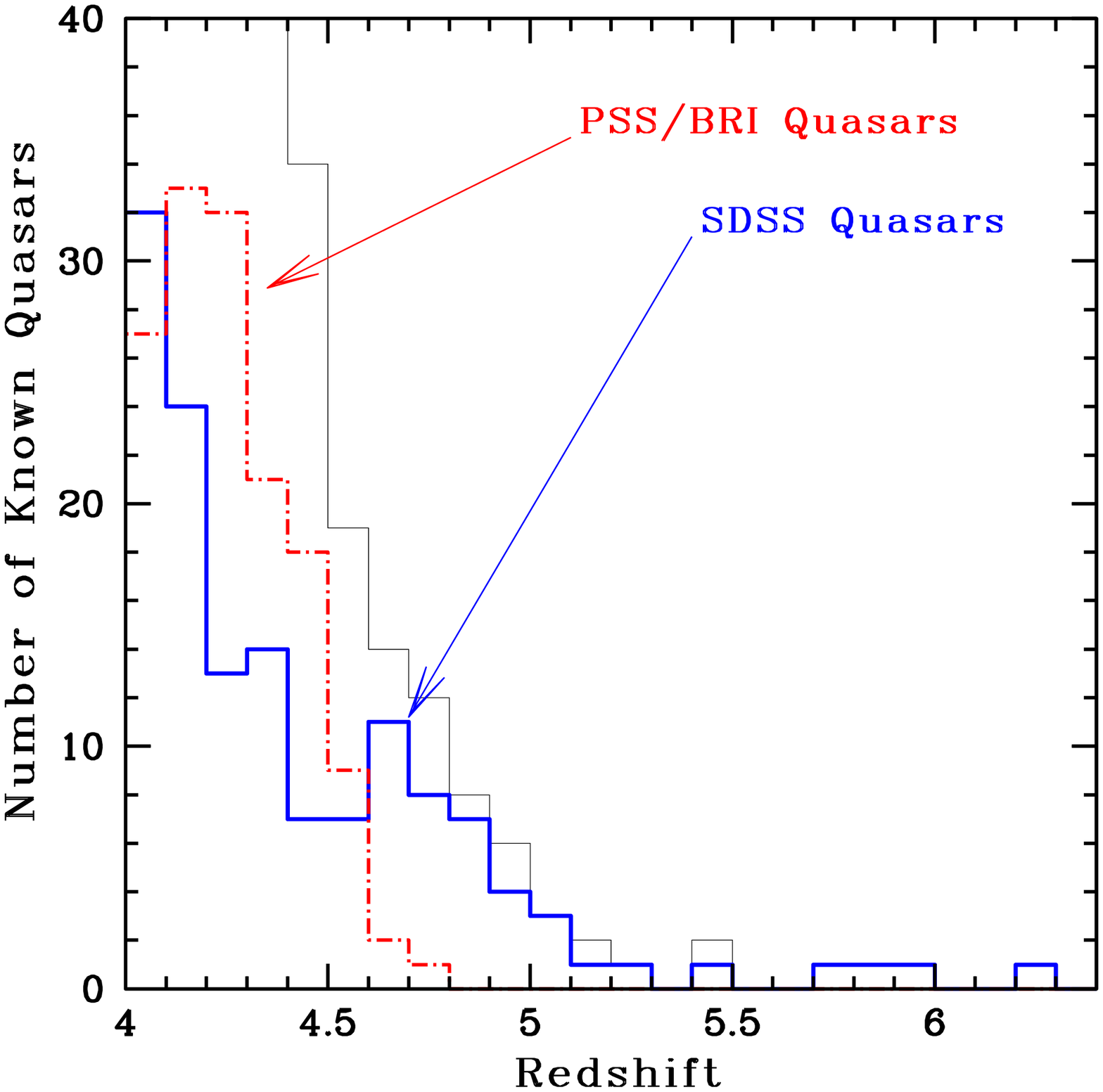,width=0.48\textwidth} 
\hspace{-0.1cm} \ \
\psfig{figure=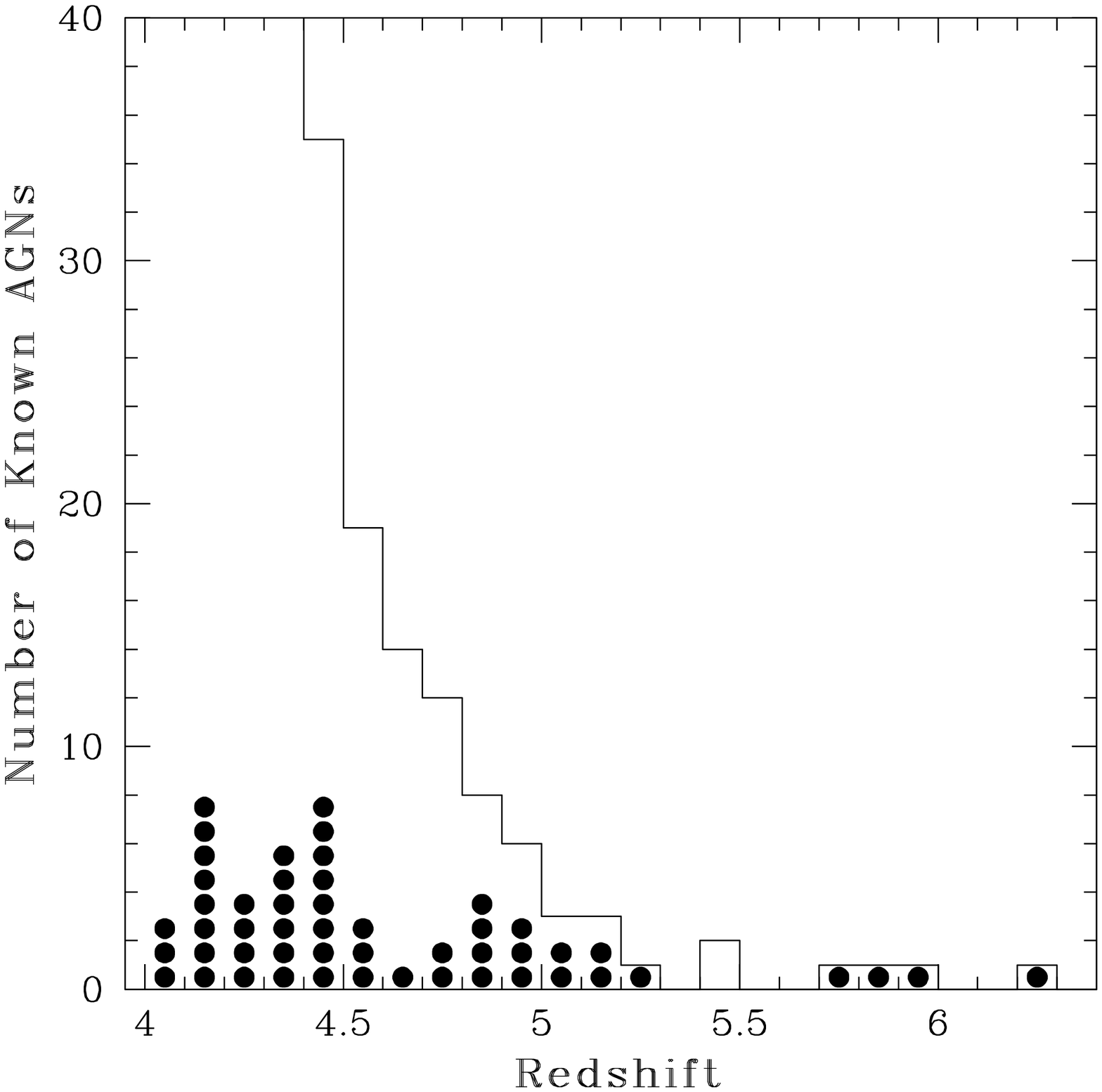,width=0.48\textwidth} 
\vskip -0.3 truecm
\caption{
(Left panel) Redshift distributions of known $z\ge4$ quasars 
(thin line), PSS/BRI quasars (dot-dashed line), 
and SDSS quasars (thick line). 
(Right panel) Redshift distribution of $z>4$ quasars with X-ray detections 
(filled circles). 
}
\label{fig1}
\end{figure}
Here we report the results obtained from \chandra\ observations of nine PSS 
quasars selected from among the optically brightest $z>4$ quasars known. 
An extensive analysis is presented in V02. 
These quasars are likely to comprise a significant fraction of the 
most luminous optically selected $z>4$ quasars over the 
forthcoming years. 
They also represent ideal targets for X-ray spectroscopy with \xmm\ and the 
next-generation of X-ray telescopes.

\section{Results}

\chandra\ and other exploratory X-ray observations have defined the typical fluxes and luminosities 
of $z>4$ quasars. As shown in Fig.~2, quasars at $z>4$ are generally faint 
\hbox{X-ray} sources, with \hbox{0.5--2~keV} fluxes 
\hbox{$<4\times10^{-14}$~\cgs}. 
The PSS quasars presented here (filled triangles in Fig.~2) 
are among the \hbox{X-ray} brightest $z>4$ quasars; they have 
\hbox{$\approx$~10--70~counts} in the observed \hbox{0.5--8~keV} band 
in a typical $\approx$~5~ks \chandra\ observation. 
The correlation between soft \hbox{X-ray} flux and \ab1450\ magnitude, 
evident in Fig.~2, is significant at 99.99\% confidence 
when the PSS/BRI/SDSS $z>4$ quasars observed by \chandra\ are taken 
into account (excluding the broad absorption-line quasars, 
which are known to suffer from intrinsic absorption{\,}\cite{g02}). 
\begin{figure}[!h]
\vskip -0.1 truecm
\centering
\epsfig{figure=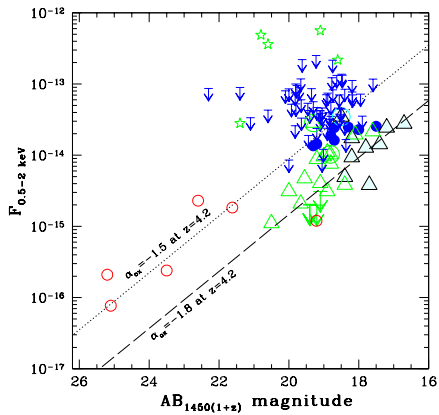,width=0.52\textwidth}
\vskip -0.3 true cm
\caption{
Observed-frame, Galactic absorption-corrected 0.5--2~keV flux 
(in units of \cgs) versus \ab1450\ magnitude for $z\ge4$ AGNs. 
The PSS quasars presented here are plotted as filled triangles, 
while the open triangles and large downward-pointing arrows indicate 
previous \chandra\ observations of $z>4$ quasars.{\,}\protect\cite{v01,b02} 
Circled triangles are radio-loud quasars. 
The quasars with \rosat\ detections or upper limits are plotted as 
filled circles and small downward-pointing arrows, 
respectively;{\,}\protect\cite{k00,v02} 
blazars are shown as open stars. 
Open circles show X-ray detected AGNs from other 
work.{\,}\protect\cite{b01a,s98,b01b,s02,ba02}
The slanted lines show $z=4.2$ loci for \aox=$-$1.5 (dotted) and \aox=$-$1.8 
(dashed).
}
\label{fig2}
\end{figure}
\vskip -0.55cm
\pn
The average \aox\ for the nine optically luminous PSS quasars is \linebreak
$\langle\alpha_{\rm ox}\rangle$=$-$1.81$\pm{0.03}$, 
considerably steeper than that obtained for samples of local quasars 
[e.g., the Bright Quasar Survey{\,}\cite{sg83} radio-quiet quasars (RQQs) 
at \hbox{$z<0.5$} have 
$\langle\alpha_{\rm ox}\rangle$=$-$1.56$\pm{0.02}${\,}\cite{blw00}]. 
A likely explanation for the range of \aox\ values found among the RQQ 
population is that 
\aox\ depends upon 2500~\AA\ rest-frame 
luminosity.{\,}\cite{v02b}
The possibility that \aox\ steepens at high redshifts 
because of the presence of \hbox{X-ray} absorption 
is ruled out by our spectral analysis of the nine PSS quasars. 
Their joint \hbox{X-ray} spectrum is well fitted by a $\Gamma=1.98\pm{0.16}$ 
power-law model (similarly to the majority of 
lower-redshift quasars{\,}\cite{rt00}) without any evidence for 
intrinsic absorption ($N_{\rm H}\simlt8.8\times10^{21}$~cm$^{-2}$ 
at 90\% confidence; see Fig.~3). 
%
\begin{figure}[!t]
\hskip -0.3cm
\centering
\epsfig{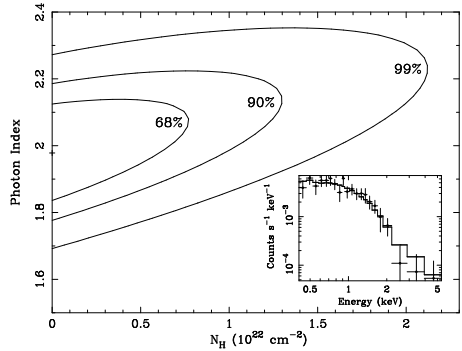}
\vskip -0.2cm
\caption{Confidence regions for the photon index 
and intrinsic column density derived from joint fitting of 
the nine PSS quasars. 
The combined quasar spectrum 
fitted with the best-fit power-law model and Galactic absorption is shown in 
the insert.}
\label{fig3}
\end{figure}

\section*{Acknowledgments}

The authors acknowledge the financial support of NASA grants 
NAS~8-01128 (GPG) and LTSA NAG5-8107 (CV, WNB), 
\chandra\ X-ray Center grant DD1-2012X (CV, WNB, DPS), 
and NSF grant AST99-00703 (DPS). 
CV also acknowledges partial support from the Italian Space Agency and MURST. 

\end{document}